\documentclass{article}

\usepackage{arxiv}

\usepackage[utf8]{inputenc} 
\usepackage[T1]{fontenc}    
\usepackage{hyperref}       
\usepackage{url}            
\usepackage{booktabs}       
\usepackage{amsfonts}       
\usepackage{nicefrac}       
\usepackage{microtype}      
\usepackage{lipsum}

\usepackage{xcolor,soul,framed} 
\colorlet{shadecolor}{yellow}
\usepackage[pdftex]{graphicx}
\graphicspath{{../pdf/}{../jpeg/}}
\DeclareGraphicsExtensions{.pdf,.jpeg,.png}
\usepackage{amsmath}
\usepackage{tabularx}
\usepackage{array}
\usepackage{mdwmath}
\usepackage{mdwtab}
\usepackage{eqparbox}
\usepackage{url}
\usepackage{cite}
\usepackage{amssymb}
\title{A Novel Method For Designing Transferable Soft Sensors And Its Application}

\author{
  Hossein Shahabadi Farahani
  \\
  APAC Research Group\\
  Department of Mechatronics\\
  K.N.Toosi University of Technology\\
  Tehran, Iran \\
  \texttt{Shahabadi.f@email.kntu.ac.ir} \\
   \And
 Alireza Fatehi \\
  APAC Research Group\\
  Department of Mechatronics\\
  K.N.Toosi University of Technology\\
  Tehran, Iran \\
  \texttt{fatehi@.kntu.ac.ir} \\

   \AND
   Alireza Nadali \\
  APAC Research Group\\
  Department of Mechatronics\\
  K.N.Toosi University of Technology\\
  Tehran, Iran \\
  \texttt{a\_nadali@email.kntu.ac.ir} \\
  
     \And
 Mahdi Aliyari Shoorehdeli \\
  APAC Research Group\\
  Department of Mechatronics\\
  K.N.Toosi University of Technology\\
  Tehran, Iran \\
  \texttt{aliyari@.kntu.ac.ir} \\

}

\begin{document}
\maketitle

\begin{abstract}
In this paper, a new approach is proposed for designing transferable soft sensors. Soft sensing is one of the significant applications of data-driven methods in the condition monitoring of plants. While hard sensors can be easily used in various plants, soft sensors are confined to the specific plant they are designed for and cannot be used in a new plant or even used in some new working conditions in the same plant. In this paper, a solution is proposed for this underlying obstacle in data-driven condition monitoring systems. Data-driven methods suffer from the fact that the distribution of the data by which the models are constructed may not be the same as the distribution of the data to which the model will be applied. This ultimately leads to the decline of models’ accuracy. We proposed a new transfer learning (TL) based regression method, called Domain Adversarial Neural Network Regression (DANN-R), and employed it for designing transferable soft sensors. We used data collected from the SCADA system of an industrial power plant to comprehensively investigate the functionality of the proposed method. The result reveals that the proposed transferable soft sensor can successfully adapt to new plants.
\end{abstract}


\section{INTRODUCTION}

{I}{ntelligent} condition monitoring plays a vital role in modern automation systems that are leading to a new industrial revolution \cite{yin2014data}. Using methods for discovering information in a huge amount of data and facilities of Internet of Things (IoT) technology, many industries tend to deploy data-driven methods in order to achieve best insights into the condition of system’s operation. In this regards, process industries have taken the advantages of data-driven condition monitoring systems by utilizing SCADA systems for collecting huge amount of data from their process operation. Soft sensing is a remarkable application of data-driven condition monitoring systems in these industries.

In general, soft sensor or virtual sensor is used to make a conclusion based upon observed process variables whenever hardware measurements are not feasible \cite{kadlec2011review, fortuna2007soft, rawat2016multi}. Actually, soft sensor is a software by which several measurement's signals are processed together in order to estimate the value of another variable of the systems. It has the advantage of being fast in responding and low in cost. Such a software can be utilized to tackle wide variety of industrial problems. It is used for reducing the cost of procurement and maintenance of hardware measurement \cite{kadlec2011review}, fault detection and diagnosis \cite{serpas2013fault, yazdani2020novel}, real-time estimation for monitoring or control \cite{etien2013modeling}, sensor validation \cite{upadhyaya1992application} and normal behaviour modeling \cite{schlechtingen2011comparative, makaremi2009abnormal}.

Basically, soft sensors can be categorized into two different types, model driven and data driven, which are also called white-box and black-box, respectively \cite{fortuna2007soft}. The former, which is out the scope of this paper, employs first principle modeling based on physical knowledge of the system. On the contrary, the latter relies on the information extracted from the system's historical data. 
    Wide arrays of machine learning techniques have been used for data-driven designed soft sensors. Most of them are based on regression problem, such as support vector regression \cite{kaneko2014adaptive}, Artificial Neural Networks (ANN) \cite{wang2019two}, gaussian regression\cite{yuan2017probabilistic}, partial least square \cite{shao2015adaptive} and so forth.
ANNs have been widely used as reliable tools for training regression models for soft sensing, since they are highly capable of capturing nonlinear dependencies of sensors data. Moreover, researchers have recently inclined to employ Deep Neural Networks (DNN) to construct better soft sensors \cite{shang2014data, yao2017deep, yan2016data, yuan2020deep, yuan2019deep}.

 Soft sensors, as data-driven models, perform well under a general assumption that the training data distribution and the test data distribution are the same \cite{ben2010theory}. 
 Unfortunately, this assumption is not satisfied in many industrial machine learning applications since data distribution is altered due to practical issues. For instance, the data collected from each plant is slightly different from the other similar types of plants \cite{yang2019intelligent}, thus, models trained using the data collected from one plant can not be directly used for prediction in another plant. Another issue is that after years of operation, plants behave differently, a phenomenon which is usually known as aging or concept drift \cite{kadlec2011review}. As a result, the model's performance is likely to decline by time. Lastly, process models are not robust to variations in working condition \cite{chen2019intelligent}. Consequently, models do not work properly when the plant meets new working conditions. From a machine learning perspective, label prediction regarding different data distributions is known as different tasks\cite{pan2009survey}. Transfer Learning (TL) algorithms aim to utilize the knowledge collected during learning a task to learn a new but related task more efficiently\cite{weiss2016survey}. TL can be used to handle problems caused by inconsistency of data distribution of industrial condition monitoring systems.
 
 TL have drawn huge attention in machine learning community since the issue of inconsistency of data distribution is a common barrier to many machine learning application, specially image processing \cite{csurka2017domain, wang2018deep}. As a result, TL methods have extensively improved by researchers in the last years. Nevertheless, little attention has been paid to the application of TL methods in improvements of the soft sensors. There exist few studies that have addressed application of TL in regression problems for condition monitoring purpose \cite{zhang2018wind, cai2019probabilistic, qureshi2017wind}, yet none of them is related to a process system. Also, some researchers have deployed TL methods in the fault detection and diagnosis of industrial systems. Yet again, rarely have these researches addressed the TL problem in process systems. They have mainly focused on fault detection of components like bearings and gearboxes in vibration systems using acceleration sensor data. For instance, \cite{shao2018highly} and \cite{han2020deep} introduced general frameworks for the fault diagnosis problem and achieved promising results in this area. To the best of authors' knowledge, only a couple of studies related to the application of TL in fault diagnosis of gas turbines, a problem which is based on classification, are available \cite{zhong2019novel, tang2019transfer}. 
 
 As motioned before, the application of TL methods for improvement of the soft sensor systems has not been much discussed so far. On the other hand, results and conclusions drawn from the studies on the fault diagnosis of vibration systems can not be generalized to process systems, since the nature of vibration systems is far from process plants. Consequently, the effectiveness of TL methods in the domain of process data and for designing soft sensors in process plants still remains unclear. In this paper, we propose a novel TL-based regression method, named Domain Adversarial Neural Networks Regression (DANN-R), and employ it for designing transferable soft sensors. The proposed method is comprehensively examined using data collected from a real-world power plant. In our studies, TL from a single gas-turbine and several gas-turbines to another gas turbines is investigated. Up to the authors knowledge, this is the first time that a transferable soft sensor is designed for a process system. Besides, the proposed method requires no labeled data from the target domain, which enables it to be employed despite of very restricted situations.

The remainder of this paper is structured as follows. The concept of TL and its formulation is elaborated in section \ref{sec_TL}. Besides, a brief review on different categories of TL approaches is provided in this section. In section \ref{sec_method}, the structure of the neural network used for DANN-R and its training algorithm is introduced. Section \ref{sec_results} is related to implementation of DANN-R for training transferable soft sensors and it consists of several parts in which the data sets and implementation results are discussed. Finally, the conclusion of this research is drawn in section \ref{sec_conclusion}.

\section{TRANSFER LEARNING}
\label{sec_TL}

Human beings are able to utilize the knowledge collected by learning a task to learn a new but related task more efficiently. The idea of TL is to actualize such a transition of knowledge in machine learning problems \cite{pan2009survey}.
Generally, the definition of TL is given in terms of  \textit{source domain} and \textit{target domain} \cite{pan2009survey}. Source domain is the domain from which the knowledge is collected and target domain is the domain to which this knowledge is applied. Generally, a domain of data is characterised by a specific data distribution. 

 The mathematical formulation of TL is as follows. The source domain is defined as $\mathcal{D}_S=\{\chi_S,P(X_S )\}$, where $ \chi_S$ is the feature space, $X_S=\{x_{S1},…,x_{Sn}\}$, $x_{Si}\in \chi_S$ is the data and $P(X_S)$ is the marginal distribution from which the source data is drawn. The corresponding ground truth of the source data is denoted by $Y_S=\{y_{S1},…,y_{Sn}\}$,  $y_{Si}\in \mathcal{Y}_S$, where $\mathcal{Y}_S$ is the output space. The assumption is that enough amount of labeled data from the source domain is available which enables training a predictive function ${\hat{f}}_S(x)$ to estimate the output $y_S$ based on the $P_S (y|x)$. Actually, ${\hat{f}}_S(.)$ is an approximation of the optimal function in the source domain $f_S(x)$.
 
 Similar to the source domain, target domain is defined as $\mathcal{D}_T=\{\chi_S,P(X_T )\}$. It is assumed that the feature space in both domains are the same, $\chi_T = \chi_S$ but their probability distributions are different, $P(X_T)\ne P(X_S)$. However, no labeled data is available in the target domain. Therefore, it is not possible to train a predictive model dedicated to the target domain which can predict corresponding labels of target data, $Y_S$, based on $P_T (y|x)$.
 On the other hand, the model trained with the source data, $\hat{f}_S (.)$, might not be an appropriate function for approximating the optimal function of the target domain, ${f}_T (x)$, $x\in \chi_T$, since the data distributions in these two domains are different. In this regards, the goal of TL methods is to find $\hat{f_T}(.)$ by using $X_S$, $X_T$ and $y_S$ so that predicts the output in the target domain more accurately compering to a model that is trained only by source data, ${\hat{f}}_S(.)$.

In our problems, $\chi$ is the space of gas-turbine’s sensors, $\mathcal{Y}$ is the space of estimated variable and $\hat{f}(.)$ is the soft sensor model. The source domain is related to the data sampled from a limited numbers of gas-turbines in a specific working condition. The target domain refers to the data sampled from gas turbine or a working condition other than the one in the source domain.


The methods and algorithms developed for TL can be categorized into three main groups \cite{pan2009survey}. 
First, instance based methods, which assign weights to source instances based on their resemblance to target data. The amount of similarity of instances is measured in a probabilistic sense. These weights are used to train a model that can make more accurate predictions regarding the target data distribution \cite{zadrozny2004learning}.

Another group of methods, which is usually referred to as parameters-based TL methods, tries to collect the knowledge from the source domain via parameters of the model trained in the source domain. It is demonstrated that the features extracted by deep layers of a neural networks are domain adaptive \cite{donahue2014decaf}. In other words, these layers can extract abstract features that are meaningful for prediction in new related domains. In this group of methods, last layers of the neural networks are usually fine-tuned by limited available target samples.

Finally, the last group of methods is representation learning based methods which is also called feature-based methods \cite{farahani2020between}. The main idea of representation learning based TL techniques is to learn a mapping that minimizes a notion of distance between domains along with the label predication risk in the source domain \cite{ben2007analysis}. For example, some studies use Maximum Mean Discrepancy (MMD) as a criterion to measure domains distances between source and target data \cite{tzeng2014deep}. Also, it is proved that the generalized error of classification between the source and target domain can be interpreted as score of divergence of the domains \cite{ben2010theory}. Based on this idea, domain adversarial training methods attempt to learn a representation in which the source and target data are indistinguishable, a goal which is achieved through the adversarial training between a feature extractor and a domain classifier \cite{ganin2016domain}. 

The idea of Domain Adversarial Training of Neural Networks (DANN) is first presented in \cite{ganin2016domain}. Afterwards, other researches inspired from this paper and introduced new TL algorithms based upon the idea of domain adversarial training of neural networks \cite{long2017deep}, \cite{hoffman2017cycada}.  Besides, domain adversarial training is successfully used in wide arrays of applications in the topic of TL. For example, domain adversarial training is employed in a fully convolutional network for medical image segmentation \cite{javanmardi2018domain}. 
Also, policies learnt in a simulation environment are transferred into real world in a robotic application using adversarial training of neural networks \cite{zhang2019adversarial}.

\section{PROPOSED METHOD} \label{sec_method}

Although TL have drawn huge attentions in the community of machine learning, the main focus of studies is on the classification problem. Consequently, few researches have paid attention to application of TL in the regression problems. In this section, we propose a neural network structure for learning transferable regression models based on DANN, called DANN-R. It is successfully employed for designing transferable soft sensors that can adapt to new plants and new working conditions.

\begin{figure*}
  \begin{center}
  \includegraphics[width=5.5in]{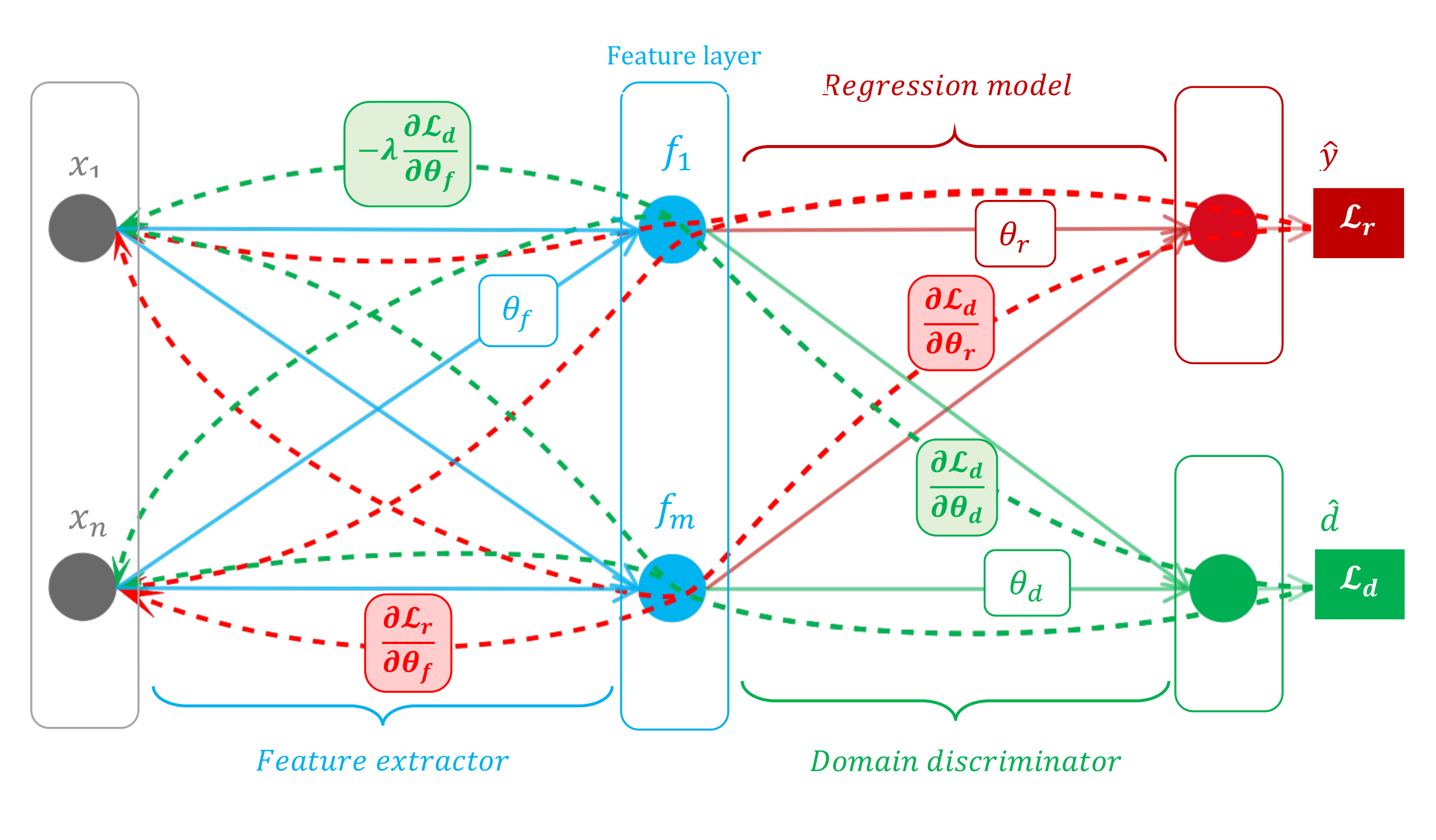}\\
  \caption{The structure of the neural networks used for DANN-R.}\label{tlreg}
  \end{center}
\end{figure*}

Fig.~\ref{tlreg} illustrates the architecture of DANN-R. This neural network consists of three major parts, feature extractor, regression model and domain discriminator. The input space is formed by an \textit{m}-dimensional input data, $\chi = \mathbb{R}^m$, which is fed into the feature extractor, $Net_f(;\theta_f)$, with the model parameters $\theta_f$. The feature extractor is a neural network that maps the input vector into an \textit{l}-dimensional feature representation, $F = \mathbb{R}^l$. Under the representation of these features, a regression model, $Net_r(;\theta_r)$, with the model parameters $\theta_r$, maps $F$ into a \textit{1}-dimensional space $\mathcal{Y}$, which represents the space of corresponding output value of the input sample, $y_i\in Y$.
Moreover, $F$ is also introduced to the domain discriminator, $Net_d(;\theta_d)$ with the model parameters $\theta_d$. $Net_d(;\theta_d)$ is a classifier that maps features $F$ into a 1-dimensional binary space $D$, which represents the space of domain label of the input sample, $d_i\in D$. In other words, the domain discriminator tries to detect whether input instances are from the source domain or the target domain. When an instance is from the source domain or target domain, the output of domain discriminators is expected to be 0 or 1, respectively.

The training procedure is so that, while the features extracted from source and target by $Net_f$ become more indistinguishable, the $Net_r$ , which is trained based on only the source data, can predict the output value of target data more accurately.
In DANN-R, such a feature extractor is found via adversarial training of $Net_f$ and $Net_d$. Actually, $Net_d$ is trained to classify the domain labels of extracted features. On the contrary, $Net_f$ is trained to extract features that can not be classified between source and target. This adversarial game between these two neural networks helps both of them to gradually learn doing their desired function during the process of training\cite{goodfellow2014generative}. Consequently, features extracted from the source and target data would be indistinguishable in terms of domain label.

The regression loss for the estimation of the output value as follow:
\begin{equation}\label{ly}
    \begin{aligned}
    \mathcal{L}_{r}^{i}\left( {{y}_{i}},{{x}_{i}};{{\theta }_{f}},{{\theta }_{r}} \right)=
    {{\left( {{(Ne{{t}_{r}}\left( Ne{{t}_{f}}\left( {{x}_{i}};{{\theta }_{f}} \right);{{\theta }_{r}} \right))}^{2}}-{{y}_{i}^{2}} \right)}^{\frac{1}{2}}}
    \end{aligned}
\end{equation}
Also, the loss of classification of domains are defined as follow:
\begin{equation}\label{ld}
    \begin{aligned}
    &\mathcal{L}_{d}^{i}\left( {{d}_{i}},{{x}_{i}};{{\theta }_{f}},{{\theta }_{d}} \right)=
    {{d}_{i}}\log \frac{1}{Ne{{t}_{d}}\left( Ne{{t}_{f}}\left( {{x}_{i}};{{\theta }_{f}} \right);{{\theta }_{y}} \right)}+\\
    &\left( 1-{{d}_{i}} \right)\log \frac{1}{1-Ne{{t}_{d}}\left( Ne{{t}_{f}}\left( {{x}_{i}};{{\theta }_{f}} \right);{{\theta }_{y}} \right)}
    \end{aligned}
\end{equation}
 Our optimization goal is to find a feature extractor so that no domain discriminator can classify the features extracted from source and target samples accurately. On the other hand, the feature extractor is needed to make an appropriate representation for the regression model in order to accurately predict the output value. To find a feature extractor with the mentioned properties, an optimization problem is proposed in which the goal is to find a saddle point that optimizes the cost function  
 
\begin{equation}\label{cost}
    \begin{aligned}
    E\left({{\theta }_{f}},{{\theta }_{r}}, {{\theta }_{d}} \right)=
    ~\frac{1}{n_S}\underset{x_i\in D_S}{\overset{}{\mathop \sum }}\,\mathcal{L}_{r}^{i}\left({{y}_{i}},{{x}_{i}};{{\theta }_{f}},{{\theta }_{r}} \right)&\\
    -
    \frac{\lambda}{n_S+n_T}\underset{x_i\in {D_S,D_T}}{\overset{}{\mathop \sum }}\,\mathcal{L}_{d}^{i}\left({d}_{i},{x}_{i}; {\theta }_{f},{\theta }_{d} \right)&
    \end{aligned}
\end{equation}

where $n_S$ and $n_T$ are the number of source and target data. As a result, optimal parameters of the networks are obtained as

\begin{equation}\label{opt}
\begin{aligned}
    \left( \hat{\theta }_f, \hat{\theta}_r \right)=~
    &{\underset {\theta_f,\theta_r} {arg~min}~E\left({\theta}_f,{\theta}_r, {\theta}_d \right)}
    \\
    \left( \hat{\theta }_d \right)=~
    &{\underset {\theta_d} {arg~min}~E\left({\theta}_f,{\theta}_r, {\theta}_d \right)}
\end{aligned}
\end{equation}

The adversarial training of the neural networks, as a technique for optimizing ANNs parameters, is employed to tackle problem (\ref{opt}). In DANN-R, $Net_f$ and $Net_d$ are trained by using gradient descent approach in an adversarial procedure, such that in each step of performing gradient descent, one is updated to minimize the cost function and the other one is updated to maximize it. As shown by Fig. \ref{tlreg}, the sign of the back propagated error from the domain discriminator is reversed after the feature layer. Meanwhile, in each step of updating model parameters, both $Net_f$ and $Net_r$ are trained to minimize the prediction error by back propagating the gradients of regression loss. Fig. \ref{tlreg} also illustrates that how the updating of $Net_f$ is influenced by gradient flows from both $\mathcal{L}_{r}$ and $\mathcal{L}_{d}$. Accordingly, The value of parameters' updates are calculated as follow:

 \begin{equation}\label{gd}
\begin{aligned}
    &{\Delta{\theta }_{f}}~=~\mu \left( \frac{\delta \mathcal{L}_{r}^{i}}{\delta {{\theta }_{f}}}-\lambda \frac{\delta \mathcal{L}_{d}^{i}}{\delta {{\theta }_{f}}} \right)\\
    &{\Delta{\theta }_{r}}~=~\mu \frac{\delta \mathcal{L}_{r}^{i}}{\delta {{\theta }_{r}}}\\
    &{\Delta{\theta_d}}~=~\mu \lambda \frac{\delta \mathcal{L}_{d}^{i}}{\delta {{\theta }_{d}}}
\end{aligned}
\end{equation}

 The hyper-parameter $\lambda$ is the feature extractor weighting parameter, which gradually decreases in each training epoch.
 
Since the parameters $\theta_f$ of $Net_f$ are updated by the gradients propagated from the $Net_d$, its updating is influenced by parameters $\theta_d$, during the training. It means that as $Net_d$ is improved during the training process, it promotes $Net_f$ to learn to extract more and more indistinguishable features. This will be discussed more in the next section.

\section{RESULTS AND DISCUSSION}
\label{sec_results}
In this section, the proposed approach for designing transferable regression models, DANN-R, is used to design soft sensors which can deal with the issue of inconsistency of data distribution in the industrial gas-turbines. These soft sensors are designed to adapt the knowledge collected from one or several plants to new plants.

            \subsection {Dataset}
We used an industrial process data set, which is collected from the SCADA system of a natural gas power plant. The power plant consists of five power units. Each of these units utilizes a Siemens\texttrademark  heavy-duty gas-turbines of class E \cite{Siemens}.

Gas-turbines experience various operation modes during their life time. Activation of a mode depends on the condition of the turbine operation. Most of these modes are related to transients, thus, they are met only during very short periods of times. Additionally, the behaviour of gas-turbines in different operation points are not the same. Consequently, data driven condition monitoring based on system's historical data during many of these modes is not meaning-full. We select SCADA data from load control and frequency control, which are dominant operation modes of the system consisting roughly 95\% of gas turbines’ data. The aim of the former mode is almost to generate the desired active power by gas-turbine, while in the latter, the control goal is to keep the turbine’s shaft rotational speed close to the desired reference. 

The data collected from different units of the power plant are used to study the ability of DANN-R algorithm for learning transferable soft sensors. In all the problems, the assumption is that target data sets are unlabeled. However, to evaluate the designed soft sensor the error between the soft sensor outputs and the actual output in the target domain is calculated.

            \subsection {Transfer learning between the plant units}
Practically, the distribution of data collected from machines or plants of the same type are different from each other. This discrepancy is derived from different maintenance events that each unit experiences, measurement settings, mechanical behavior, and so forth. In this part, the aim is to evaluate the capability of DANN-R for TL between different machines. The source and target data are collected from the gas-turbines of different units of the power plant. Therefore, all environmental condition, like the temperature, humidity and the plant site altitude are the same.

In this part, we design soft sensors that predict the value of the active power. The input variables of the soft sensor models are introduced in Table \ref{set1}. This set of variables are selected according to the performance analysis of the gas-turbines \cite{hanachi2018performance}, \cite{bartolini2011application}. 
Three single layer neural networks with proper dimensions are selected for the feature extractor, domain discriminator and regression model. 
This is different from the feature extractor in original DANN [14] used for classification where ReLU activation function is used. 
We used sigmoid activation function in feature extractor, since we find it more functional for capturing non-linear relations in data sets when the depth of networks in not high.
The feature extractor consists of 60 neurons with sigmoid activation function. The regression model and domain discriminator are logistic regression layer and regression layer, respectively. 

\begin{figure}
  \begin{center}
  \includegraphics[width=3.5in]{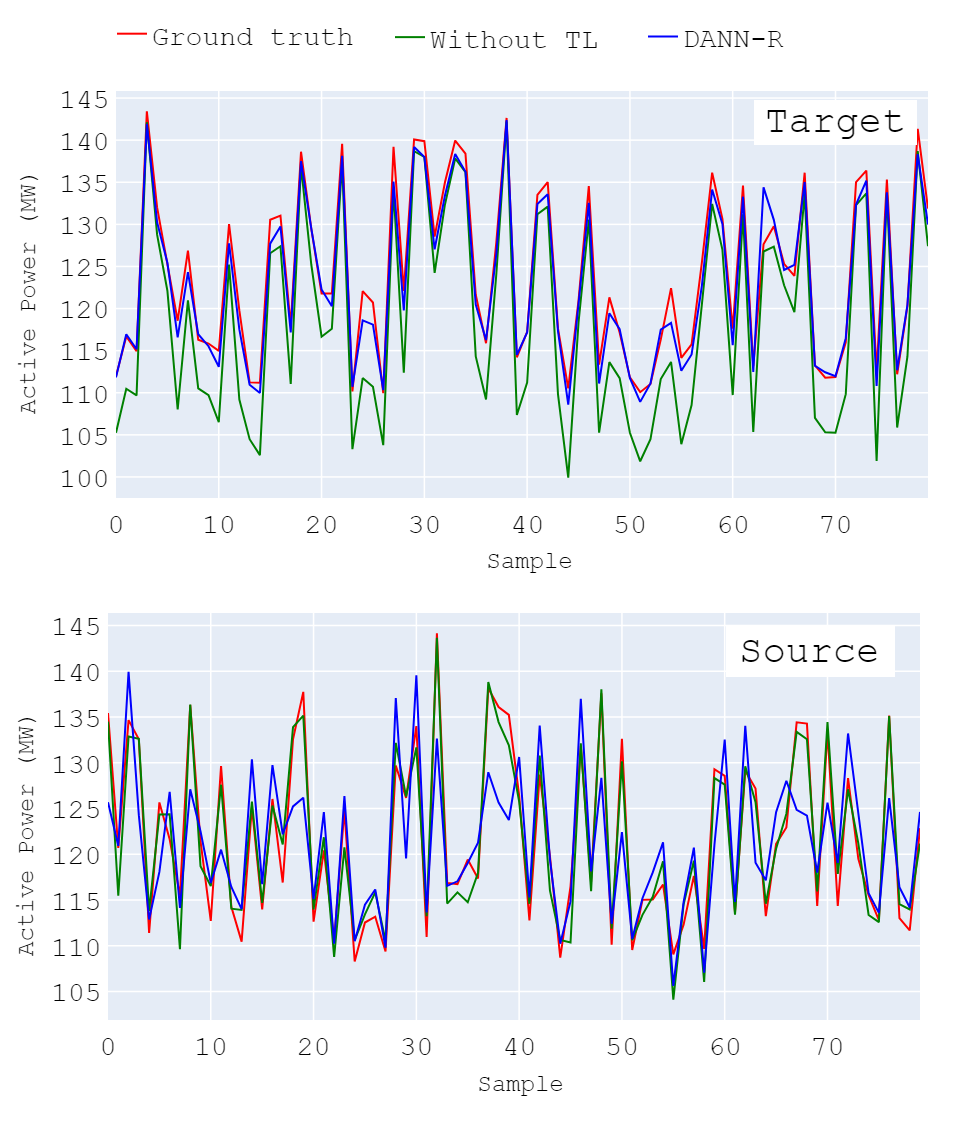}\\
  \caption{Transfer learning from a gas-turbine to another gas-turbine.}\label{m2m}
  \end{center}
\end{figure}

Fig. \ref{m2m} depicts the results of applying DANN-R for designing a transferable soft sensor in the case of TL between two different plants.
The figure includes the plots related to the prediction of models in the both source and target domains.
The real value of the active power, the prediction of the model trained without TL and the prediction of our transferable soft sensor are shown respectively by red, green and blue plots. The model trained without TL, means the model that is trained by using only source data.
In the target domain, the green plot is unable to properly follow the ground truth. On the other hand, it can be seen that the prediction of the proposed DANN-R transferable soft sensor, i.e., the blue plot, provides far more accurate estimation for the real value of the active power.

\begin{table}
\begin{center}
\caption{Input and output of models studied in the case of transfer learning between different machines.}\label{set1}
\begin{tabular}{l|l}
\textbf{Input Sensors}                                                             & \textbf{Estimated Sensor} \\ \hline
\begin{tabular}[c]{@{}l@{}}Ambient temperature\\ Ambient humidity\\ IGV angle\\ Fuel flow\end{tabular} & Active power             
\end{tabular}
\end{center}
\end{table}

Fig. \ref{m2m} also shows the performance of trained models in the source domain. Usually, in the source domain, the prediction of the models trained without TL is better compered to the models trained using DANN-R. 
Indeed, using DANN-R for designing transferable soft sensors lowers the performance of models in the source domain, which is caused by the multi-objective training procedure of DANN-R. The term related to the adversarial training of domains in (\ref{cost}), in a sense, interfere with training of the model for regression in the source domain. Actually, the adversarial game between the feature extractor and the domain discriminator prevents the feature extractor from providing the best possible features for regression in the source domain.
However, in TL problems, the accuracy of models in the source domain is not a concern because the main focus is to enhance the models performance in the target domain, where the intended model operates.

Table \ref{m2m_single} and Table \ref{m2m_multiple} provide quantitative results of implementations in this scenario. 
 We have defined the \textit{transfer ratio} score in order to evaluate the performance of transferable soft sensors. It is calculated as follow: 
\begin{equation}\label{tlratio}
    Transfer\: Ratio = \frac{Target\:   MSE\: without\:   TL}{Target\:   MSE\: using\:   TL}
\end{equation}
 This score by some means shows that how the transferable soft sensor can enhance the estimation of the output variable.
 With industrial applications in mind, \textit{transfer ratio} in both tables are appealing.

\begin{table*}[]
\caption{MSE of soft sensors prediction in cases of transfer learning from one machine to another machine}\label{m2m_single}
\begin{center}
\begin{tabular}{|c|c|c|c|c|c|c|}
\hline
\textbf{Target}    & \textbf{Source}   & \textbf{\rotatebox[origin=c]{90}{Source MSE Without TL}} & \textbf{\rotatebox[origin=c]{90}{ Source MSE Using TL }} & \textbf{\rotatebox[origin=c]{90}{ Target MSE without TL }} & \textbf{\rotatebox[origin=c]{90}{ Target MSE Using TL }} & \textbf{\rotatebox[origin=c]{90}{ transfer ratio }} \\ \hline
Unit 1             & Another turbine   & 0.0016                         & 0.0013                       & 0.0262                         & 0.0219                       &   1.20                  \\ \hline
Unit 2             & Another turbine   & 0.0017                         & 0.0059                       & 0.0061                         & 0.0022                       &   2.77                  \\ \hline
Unit 3             & Another turbine   & 0.0059                         & 0.0030                       & 0.0235                         & 0.0116                       &   2.03                  \\ \hline
Unit 4             & Another turbine   & 0.0054                         & 0.0062                       & 0.0112                         & 0.0031                       &   3.61                  \\ \hline
Unit 5             & Another turbine   & 0.0015                         & 0.0072                       & 0.0314                         & 0.0062                       & 5.06                  \\ \hline
\multicolumn{2}{|c|}{\textbf{Average}} & 0.0032                         & 0.0047                       & 0.0197                         & 0.0090                       & \textbf{2.93}         \\ \hline
\end{tabular}
\end{center}
\end{table*}

\begin{table*}[]
\caption{MSE of soft sensors prediction in cases of transfer learning from multiple machines to one machine}\label{m2m_multiple}
\begin{center}
\begin{tabular}{|c|c|c|c|c|c|c|}
\hline
\textbf{Target}    & \textbf{Source}   & \textbf{\rotatebox[origin=c]{90}{Source MSE Without TL}} & \textbf{\rotatebox[origin=c]{90}{ Source MSE Using TL }} & \textbf{\rotatebox[origin=c]{90}{ Target MSE without TL }} & \textbf{\rotatebox[origin=c]{90}{ Target MSE Using TL }} & \textbf{\rotatebox[origin=c]{90}{ transfer ratio }} \\ \hline
Unit 1           & All other turbines  & 0.0019                      & 0.0018                    & 0.0144                      & 0.0136                    & 1.06             \\ \hline
Unit 2           & All other turbines  & 0.0032                      & 0.0026                    & 0.0111                      & 0.0071                   & 1.56             \\ \hline
Unit 3           & All other turbines  & 0.0047                     & 0.0044                   & 0.0025                      & 0.0008                   & 3.13                \\ \hline
Unit 4           & All other turbines  & 0.0043                     & 0.0041                   & 0.0017                      & 0.0014                   & 1.21             \\ \hline
Unit 5           & All other turbines  & 0.0063                     & 0.0064                    & 0.0038                     & 0.0018                   & 2.11               \\ \hline
\multicolumn{2}{|c|}{\textbf{Average}} & 0.0041                     & 0.0038                   & 0.0067                    & 0.0049                  & \textbf{1.81}    \\ \hline
\end{tabular}
\end{center}
\end{table*}

In both Table \ref{m2m_single} and Table \ref{m2m_multiple}, target domain is related to a single gas-turbine. Table \ref{m2m_single} presents implementations results in which only one other gas-turbine is selected as the source domain, while Table \ref{m2m_multiple} presents the results of implementations in which the source domain includes all other gas-turbines of the power plant.

The average of the MSE in the target domain without TL in the Table \ref{m2m_single} is in average about 5.18 times higher than that of Table \ref{m2m_multiple}, but source MSE without TL in this table is in average 1.25 times lower. This results are not far from expectations since each source domain in Table \ref{m2m_multiple} includes data collected from other four turbines. In other words, data distribution of source domains in Table \ref{m2m_multiple} are richer, thus, these data sets have a better generalization that enables the models trained in the source domain, even without TL, to make relatively accurate predictions on the target domain.
On the other hand, it is more difficult to train an accurate model for regression in the source domains of the experiments in Table \ref{m2m_multiple} because of high diversity of the data, which results in increase of the MSE in source domain.

 Comparing the MSE of the target domain in Table \ref{m2m_single} and Table \ref{m2m_multiple} with and without TL reveals that the proposed method can enhance the generalization of the trained model to the target gas-turbine by remarkable ratios and decrease the MSE of models in the target domain by 2.93 and 1.81 times on average, respectively. On the other hand, the improvement of Table \ref{m2m_multiple} suggests that even in many cases that training set consists of rich data sets and it may be believed that the generalization of the training data is high, still DANN-R is likely to considerably improve the performance of the models in the desired domain in which the model will be used.

The results in Table \ref{m2m_single} and Table \ref{m2m_multiple} also indicates that the transfer learning degrades the performance of the regression model in source domain. As mentioned before, the reason for this fact is that during the training of models using DANN-R, along with predicting the output variable in the source domain, the models are also trained to source and target domains. Therefore, they are not optimally trained to perform the regression task in the source domain.
\section{CONCLUSION}
\label{sec_conclusion}

In this paper, we successfully contributed to finding a solution for a crucial obstacle in data-driven condition monitoring of process systems which is caused by the issue of inconsistency of data distribution. We proposed a Transfer Learning (TL) based regression method for designing transferable soft sensors, Adversarial Neural Networks Regression (DANN-R), and extensively examined it by data collected from SCADA system of a power-plant.

We demonstrated that by using our method it is possible to extract knowledge from historical data of a gas turbine in a specific working condition and transfer it to another gas-turbine. The problem to which we tackled in this paper is likely to challenge condition monitoring systems in industrial practices of digitization. We demonstrated that a model trained with a rich data set ,which consists of samples from several gas-turbines fleet, has a great generalization by itself, yet still our method decrease the MSE of predictions on the target domain by 1.82 times on average.

The functionality of TL methods for soft sensor problems in process systems has been uncertain so far. Providing an approach for designing transferable sensors, this paper reveals that TL can dramatically enhance the performance of models in these problems. By using our transferable soft sensor, it is possible to predict the value of sensors that are defective or not installed in a power plants via the knowledge transferred from other gas turbine fleets. For instance, Lower Heating Value (LHV) sensor \cite{Bhatia2012}, which the hardware sensor is hard to be maintained and expensive to operate. Furthermore, there are some sensors that are installed in system only during limited periods of time, for example during Performance Guarantee Test (Commonly known as PG Test) whose values are very useful for condition monitoring purposes. A model trained with data gathered during such a limited period of time might not be able to make accurate predictions in all working conditions. Again in such cases, our transferable soft sensor can be useful.

\bibliographystyle{IEEEtran}
\bibliography{IEEEabrv,main}

\vfill

\end{document}